\documentclass[aps,superscriptaddress,showpacs,nofootinbib,twocolumn]{revtex4}

\input{epsf}

\begin{document}

\title{Unusual Transition Patterns in Bose-Einstein Condensation}

\author{Marcus B. Pinto}
\affiliation{Departamento de F\'{\i}sica,
Universidade Federal de Santa Catarina,
88040-900 Florian\'{o}polis, SC, Brazil}
\affiliation{Laboratoire de Physique Math\'{e}matique et 
Th\'{e}orique - CNRS - UMR 5825 Universit\'{e} Montpellier II, France}

\author{Rudnei O. Ramos}
\affiliation{Departamento de F\'{\i}sica Te\'orica,
Universidade do Estado do Rio de Janeiro,
20550-013 Rio de Janeiro, RJ, Brazil}

\author{Frederico F. de Souza Cruz}
\affiliation{Departamento de F\'{\i}sica,
Universidade Federal de Santa Catarina,
88040-900 Florian\'{o}polis, SC, Brazil}

\begin{abstract}

We analyze the possible transition patterns exhibited by an effective
non-relativistic field model describing interacting binary homogeneous
dilute Bose gases whose overall potential is repulsive. We evaluate the
temperature dependence of all couplings and show that at intermediate
temperatures the crossed interaction, which is allowed to be attractive,
dominates, leading to smooth re-entrant phases. At higher temperatures
this interaction suffers a sudden sign inversion leading to an abrupt
discontinuous transition back to the normal gas phase. This situation
may suggest an alternative way to observe collapsing and exploding
condensates.  
Our results also suggest that such
binary systems may offer the possibility of observing Bose-Einstein
condensation at higher critical temperatures.

\end{abstract}

\pacs{11.10.Wx, 98.80.Cq, 03.75.Fi, 05.30.Jp}

\maketitle

The study of symmetry breaking (SB) and symmetry restoration (SR)
mechanisms have proven to be extremely useful in the analysis of
phenomena related to phase transitions in almost all branches of
physics. Some topics of current interest which make extensive use of
SB/SR mechanisms are topological defects formation in cosmology, the
Higgs-Kibble mechanism in the standard model of elementary particles and
the Bose-Einstein condensation (BEC) in condensed matter physics. An
almost general rule that arises from those studies is that a symmetry
which is broken at zero temperature should get restored as the
temperature increases. Examples range from the traditional ferromagnet
to the more up to date chiral symmetry breaking/restoration in QCD, with
the transition pattern being the simplest one of going from the broken
phase to the symmetric one. 

A counter-intuitive example may happen in multi-field scalar models, as
first noticed by Weinberg \cite{weinberg}. Considering an $O(N)\times
O(N)$ invariant relativistic model, with two types of scalar fields and
different types of self and crossed interactions, Weinberg has shown
that it is possible for the crossed coupling constant to be negative,
while the model is still bounded from below, leading, for some parameter
values, to an enhanced symmetry breaking effect at high temperatures.
This would predict that a symmetry which is broken at $T=0$ may not get
restored at high temperatures, a phenomenon known as symmetry
nonrestoration (SNR), or, in the opposite case, a symmetry that is
unbroken at $T=0$ would become broken at high temperatures, a case
called inverse symmetry breaking (ISB). Since then the model has been
re-investigated by many other authors using a variety of different
methods, both perturbative and nonperturbative analytical and numerical
methods, giving further support to the idea of SNR/ISB. A recent review
\cite {borut} lists most applications and gives an introduction to the
subject, discussing other contexts in which SNR/ISB can take place in
connection with cosmology and condensed matter physics.

Two of the present authors have also treated the problem
nonperturbatively taking full account of the cumbersome two-loop
contributions \cite {MR1}. The results obtained in Ref. \cite{MR1} were
shown to be in good agreement with those of Ref. \cite {roos}, where the
SNR/ISB phenomena were studied using the Wilson Renormalization Group
(WRG) and the explicit running of the (temperature) dependent coupling
constants has been taken into account, showing that in fact the strength
of all couplings increase with the temperature, enhancing SNR/ISB. This
result completely rules out a possible decrease of the strength of the
negative crossed coupling that would lead to the eventual SR at higher
temperatures. These interesting results from finite temperature quantum
field theory raise important questions regarding a possible experimental
observation of those phenomena in current laboratory experiments.

The recent experimental achievement of BEC and its further improvements
has opened the interesting possibility of probing and studying finite
temperature quantum field models and methods, currently used in
cosmology and particle physics, in the laboratory. One of the remarkable
things about BEC of dilute atomic gases is the possibility of
adjusting several experimental features by fine-tuning the parameters
with a high level of control and accuracy (for recent reviews see Ref.
\cite{becreview}). We could, for instance, envisage the possibility of
investigating the SNR/ISB phenomena using a system composed by a mixture
of coupled atomic gases, like the ones recently produced \cite {myatt}
in which one has the same chemical element in two different hyperfine
states and that may be treated as ``effectively distinguishable", or
just consider the mixing of two different mono-atomic Bose gases. 

Here, we shall do a theoretical investigation of the possible transition
patterns followed by this type of system paying special attention to the
SNR issue since one expects, on basic grounds, that as the temperature
increases the atoms will leave the condensate phase  going  to
the symmetric phase of a normal gas. As a byproduct we shall see how the
crossed couplings shift the critical temperatures with respect to those
obtained for uncoupled gases, creating the possibility of obtaining BEC
at higher temperatures.

The model we consider is similar to the ones used in other theoretical
studies of homogeneous dilute coupled Bose gases \cite{coupled}, that
consists of a hard core sphere gas model described by non-relativistic
interacting (complex) scalar fields, with an overall repulsive
potential. This system can be described by the following $U_\psi(1)
\times U_\phi(1)$ invariant finite temperature Euclidean spacetime
action

\begin{eqnarray}
\lefteqn{ \!\!\!\!\! \!\!\!\! S_E (\beta) \! = \!\! \int_0^\beta \! d \tau \! 
\int \! d^3 x \! \left[\!
\psi^* \! \left( \frac{\partial}{\partial \tau} \! - \!
\frac{\nabla^2 }{2 m_\psi} \!  - \! \mu_\psi \right) \psi \! + \!
\frac {g_{\psi}}{4}(\psi^*\psi)^2
\right. }  \nonumber \\
&&  \!\!\!\!\! \!\!\!\!\!\!\!\!\! + \left. \phi^* \!\! 
\left( \! \frac{\partial}{\partial \tau} \!-
\! \frac{\nabla^2 }{2 m_\phi} \! - \! \mu_\phi \! \right) \! \phi \! 
+ \! \frac {g_{\phi}}{4}(\phi^*\phi)^2 \!
+ \! g  (\psi^*\psi)(\phi^*\phi) \right] ,
\label{action}
\end{eqnarray}

\noindent
where, in natural unities, $T=1/\beta$. The associated chemical
potentials are represented by $\mu_i$ ($i=\psi$ or $\phi$) while $m_i$
represent the masses. {}For the hard core sphere self-interactions we
take the phenomenological coupling constants as being the ones normally
used in the absence of crossed interactions and which are valid in the
dilute gas approximation \cite{becreview}. In terms of the corresponding
$s$-wave scattering lengths $a_i$ they can be written as $g_i = 8 \pi
a_i/m_i$. To make contact with the analogous potential used in the
prototype relativistic models for SNR, we take the overall potential as
being repulsive, bounded from below. This requirement imposes the
constraint condition $g_{\psi} > 0$, $g_{\phi} > 0$ and $g_{\psi}
g_{\phi} > 4 g^2$.

At $T=0$, in the absence of crossed interactions and for $ \mu_i >0$ the
classical potential of Eq. (\ref {action}) exactly reproduces the case
of an uncoupled gas with symmetry broken ground states representing
``condensates" $|\psi| = (2 \mu_{\psi}/g_{\psi})^{1/2}$ and $|\phi| = (2
\mu_{\phi}/g_{\phi})^{1/2}$. At the same time, the case $ \mu_{\psi }
<0$ and $ \mu_{\phi} < 0$ corresponds to the normal symmetric phases
without any condensates, $|\psi| = 0$ and $|\phi| = 0$. Our aim is to
investigate, for non-negligible binary interactions, how the inclusion of
temperature corrections can alter this picture. This can be achieved by
computing the temperature dependent effective chemical potentials (see
for instance Ref. \cite {norway} for an analogous analysis for the
mono-atomic gas) defined as a solution of the gap equation $\mu_i(T)=
\mu_i(0) + \Sigma_i^T({\bf p})$, where $\Sigma_i^T({\bf p})$ is the
field temperature dependent self-energy. 

The phase structure of the model is then given by the sign of $\mu_i(T)$
at a given temperature. One has $\mu_i(T<T_c^i)>0$ in the broken
condensate phase and $\mu_i(T>T_c^i)<0$ in the symmetric normal-gas
phase. At the same time the Hugenholtz-Pines theorem imposes
$\mu_i(T=T_c^i)=0$. Our next step is to evaluate the thermal
self-energies which we shall do in a nonperturbative self-consistent
fashion so as to avoid any potential problems associated with the usual
perturbative calculation (see, {\it e.g.}, Ref. \cite{MR1} for a
discussion in relativistic field theory). One can perform a one-loop
self-consistent resummation by using the effective dressed propagator
$D_{i,i^*} ({\bf p}, \omega_n) = [- i \omega_n + \omega_i]^{-1} $ where
$\omega_n = 2 \pi n/\beta$ are the bosonic Matsubara frequencies and
$\omega_i= {\bf p}^2/(2m_i) - \mu_i(T)$. One gets

\begin{equation}
\!\!\Sigma_i^T({\bf p})\! =\!
- \frac {1}{\beta}\sum_{n} 
\int \frac {d^3 {\bf p}}{(2 \pi)^3}
\left[\!
\frac {g_i(0)}{-i \omega_n + \omega_i}
\!+\!\frac {g(0) }{-i \omega_n + \omega_j} \! \right],
\label{sigma}
\end{equation}
where $g_i(0)$ and $g(0)$ indicate the bare, zero temperature,
coupling constants.
The sum in Eq. (\ref{sigma}) can be easily performed and the resulting
momentum integrals lead to well known Bose integrals \cite{pathria}.
{}For simplicity, in the following we consider atoms (fields) with
approximately the same mass, $m_\psi \simeq m_\phi=m$. One  obtains
two one-loop self-consistent coupled equations given by

\begin{eqnarray}
\lefteqn{\!\!\!\!\!\!\!\!} \mu_i(T) &=&   \mu_i(0) -
g_i(0)\left ( \frac { m}{2 \pi \beta} \right )^{3/2}
{\rm Li}_{3/2}[\exp (\beta \mu_i(T))]
\nonumber \\     
&-&  g(0)  \left ( \frac {m}{2 \pi \beta} \right )^{3/2}
{\rm Li}_{3/2} [\exp (\beta \mu_j(T))] \;,
\label{mueff}
\end{eqnarray}
where ${\rm Li}_{n}(z)$ is the polylogarithmic function. As shown
below, $\mu_i(0) << T_c$ for realistic parameter values. We shall also
see that $\mu_i(T)$ quickly decreases (in the SR case) or remains
approximately equal to $\mu_i(0)$ (in the SNR case) with increasing $T$
so that one may safely consider the high temperature approximation
$\mu_i(T)/T << 1$ by taking ${\rm Li}_{3/2} [\exp (\beta \mu_j(T))]
\sim \zeta(3/2)$. One then obtains the critical temperatures

\begin{equation}
T_c^i = \left (\frac {2\pi}{m} \right ) 
\left \{ \frac {\mu_i(0)}{[g_i(0) + g(0)]\zeta(3/2)} \right \}^{2/3}\;.
\label{tc}
\end{equation}
Eq. (\ref{tc}) displays some unusual effects due to the presence of
crossed interactions. There are three interesting cases which depend on
the sign and magnitude of $g(0)$. Taking $g(0)>0$ one observes a shift
in the critical temperatures indicating that the BEC/normal-gas
transition occurs at lower temperatures compared to the usual
mono-atomic case ($g(0)=0$). If $g(0) <0$ but $|g(0)|< g_i(0)$ then the
transition occurs at higher temperatures. Despite these important
quantitative differences symmetry restoration does take place in both
cases. Now consider the case where $g(0)<0$ but $|g(0)| > g_\phi(0)$ (in
this case the boundness condition assures that $|g(0)| < g_\psi(0)$).
{}For this situation something unexpected occurs concerning the $\phi$
field since Eq. (\ref {tc}) does not allow for a finite, positive real
critical temperature value. This is a manifestation of SNR within our
two-field model and is analogous to what is seen in the relativistic case. 
At the same
time the field $\psi$ suffers the expected phase transition at a higher
$T_c$ compared to the $g(0)=0$ case. Obviously, which field will suffer
SNR depends on our initial choice of couplings. 

{}For the following analysis let us set the parameters by choosing
representative values for a mixture of gases such as $^{85}{\rm Rb}$ and
$^{87}{\rm Rb}$. Some realistic values are $m=86 {\rm GeV}$ and $a_\psi
= 2.5 \times 10^{-2} ({\rm eV})^{-1}$, corresponding to the $s$-wave
scattering length of $^{87}{\rm Rb}$, which fix the coupling $g_\psi(0)=
7.3 ({\rm MeV})^{-2}$. Setting $\mu_\psi(0)= 5.67 {\rm peV}$ the value
$T_c\simeq 32.5 {\rm peV} = 280 {\rm nK}$ is reproduced in the usual
$g=0$ case. Throughout this Letter we set $\mu_\phi(0)=\mu_\psi(0)$,
keep $g_\psi(0)$ fixed, while considering $g_\phi(0)$ and $g(0)$ as
tunable parameters. In principle, this can be experimentally achieved by
appropriately setting magnetic fields close to a {}Feshbach resonance as
described in a recent application to $^{85}{\rm Rb}$ \cite {magn}. As an
illustration of the relevant phenomena we want to describe, we consider
three possible sets of numerical values for $g_\phi(0)$ and $g(0)$,
given, in unities of ${\rm MeV}^{-2}$ by: I) $g_\phi(0)= 5.6 $ and
$g(0)= 2.3 $; II) $g_\phi(0)= 5.6 $ and $g(0)= - 2.3$ and III)
$g_\phi(0)= 0.56 $ and $g(0)= - 0.84 $. Based on the general results
given by Eqs. (\ref{mueff}) and (\ref{tc}) we then have the usual
transition patterns for both $\psi$ and $\phi$ for sets I and II where
$\mu_\psi(T)$ and $\mu_\phi (T)$ go from positive to negative values,
with transition temperatures that can be easily derived from Eq.
(\ref{tc}). {}For set III, the field $\psi$ exhibits the usual transition
from the condensed to the normal gas phase as the temperature is
increased from zero. However, we have $\mu_\phi(T) >0$ for arbitrarily
large temperatures, which corresponds to SNR. This result is however
misleading, as we demonstrate below. 

As a general result from the renormalization group, we expect that
all coupling constants should run with the temperature \cite {roos}.
{}For instance, if $g_i(T)$ decreases faster than $g(T)$, or on the other
way around, if
the crossed coupling decreases faster than the self-couplings,
the simple analysis performed above can change drastically.
As noted in the introduction, those substantial qualitative changes were not
observed in the relativistic case.
However, a basic difference, in between the relativistic and non-relativistic
model studied here, refers to the type of four-point
functions allowed by each model. In fact, the contributions considered
in the relativistic calculations include the $t$ and $u$ scattering
channels as well as the $s$-channel contributions. At the same time,
elastic and inelastic collisions are allowed whereas for the present
case only elastic, $s$-channel contributions can be considered since our
effective model represents a system of hard core spheres. Keeping
these facts in mind and performing a nonperturbative one-loop
calculation in terms of temperature dependent vertices and effective
propagators one obtains

\begin{equation}
g(T) =
g(0)\left[1 + g(0) B_{i,j}({\bf k})\right]^{-1},
\end{equation}
where $B_{i,j}({\bf k})$ represents the bubble contribution,

\begin{equation}
B_{i,j}({\bf k}) = s_{i,j} \; \beta \int_0^1 d \alpha \int \frac {d^3 {\bf q}}{(2 \pi)^3}
 \frac {\exp (\beta \Omega_{i,j})}
{\left [ \exp (\beta \Omega_{i,j})-1
\right ]^2 } \;.
\label{Bij 2}
\end{equation}
The quantity $s_{i,j}$ is a symmetry factor, $s_{i,j}=2$ for $i\neq j$
and $s_{i,j}=5/2$ for $i =j$, while $\Omega_{i,j}$
is given by

\begin{equation}
\Omega_{i,j}\!=\!  \frac {{\bf q}^2
\!+\!  \alpha  (1-\alpha)  {\bf k}^2 }{2m}
\!-\!   \alpha \mu_i(T) \!-\! (1-\alpha) \mu_j (T)  \; ,
\end{equation}
where ${\bf k}$ represents the $s$-channel incoming momentum and
$\alpha$ is a {}Feynman parameter introduced to merge the two internal
bubble propagators in the derivation of Eq. (\ref{Bij 2}). The ${\bf q}$
integral in Eq. (\ref{Bij 2}) is again performed with the help of the
Bose integrals and one obtains

\begin{equation}
B_{i,j}({\bf k})\!=\! s_{i,j} \beta \left( 
\frac { m}{2 \pi \beta} \right)^{3/2} \!
\int_0^1 \! d \alpha \;
{\rm Li}_{1/2}[\exp (-\beta \chi_{i,j} )],
\label{Bij 3}
\end{equation}
where 
$\chi_{i,j} \!=\! \left [ \alpha (1\!-\!\alpha) {\bf k}^2/(2m)
-\mu_i(T)\alpha \!-\! \mu_j(T)(1-\alpha) \right ]$. Using the expansion
{${\rm Li}_{1/2}[\exp (-z)]=1.77/z^{1/2} -1.46 + 0.2082 z - 0.0128 z^2 +
{\cal O}(z^3)\;\;$} (see Ref. \cite {pathria}) and setting ${\bf
k}^2/2m=3T$ (the average incoming two particle kinetic energy), the
$\alpha$ integral in Eq. (\ref{Bij 3}) can be performed. In the high
temperature limit, one obtains

\begin{equation}
g(T)  \simeq g(0)\left[1 + \gamma_{i,j} \; g(0)\; T^{1/2}
\left ( \frac { m}{2 \pi } \right )^{3/2} \right]^{-1} \;,
\label{gt}
\end{equation}
where $\gamma_{i,j} = 3.7012$. The equations for the couplings
representing the self-interactions, $g_\psi (T)$ and $g_\phi(T)$, have
exactly the same structure and can be obtained by replacing $g$ with
$g_\psi$ or $g_\phi$ in Eq. (\ref{gt}) and $\gamma_{i,j}$ by
$\gamma_{i,i}= 4.6265$. These results show that $g_\psi(T)$ and
$g_\phi(T)$ always decrease, approaching zero monotonically at high
temperatures since $g_\psi(0)$ and $g_\phi(0)$ are positive quantities.
The same happens for $g(T)$, if $g(0)$ is positive. However, for $g(0)
<0$, it can be easily checked that $g_\psi (T)$ and $g_\phi(T)$ decrease
faster than $g(T)$. Consequently the transition picture described above
will be shown to change in an unexpected way. In this case, Eq.
(\ref{gt}) shows that $g(T)$ is initially negative with an absolute
which increases up to a certain critical inversion temperature given by
$T_g= ( 2 \pi/m )^3 [3.7012 \; g(0)]^{-2}$, where it develops a pole.
Then, for $T>T_g$, $g(T)$ suddenly becomes positive which means that,
contrary to the relativistic case, one always has symmetry restoration
at high temperatures in the nonrelativistic model of hard core spheres.
It is important to note that this pole in Eq. (\ref{gt}) has a
completely different origin from the usual Landau pole found in the
relativistic $\phi^4$ theory. Here the pole signals a phase transition
and if the nonperturbative self-consistent loop expansion used here
converges, as generically expected, a next order calculation will only
lead to a small change in the value of $T_g$. 

It is instructive to look in more detail at sets II and III where $g(0)$
is negative. The results for $\mu_i(T)$ are shown in {}Fig. 1. Contrary
to the previous analysis, with bare (temperature independent) coupling
constants, we now have a completely unexpected behavior for both fields
for some parameters values, as the ones exemplified by sets II and III.
{}For set III, {}Fig. 1 shows that $\phi$ displays the typical SNR
behavior while $\psi$ undergoes SR. The same figure shows that, for set
II, both fields observe SR. In the three SR cases above, the transition
from the BEC phase to the normal-gas phase is observed at critical
temperatures between $380 {\rm nK}$ and $900 {\rm nK}$. Then, as soon as
the negative $g(T)$ reaches a magnitude which is larger than that of
$g_i(T)$, both type of atoms return to the BEC phase in a re-entrant
transition. We note that, for set III, the field $\psi$ also goes through
a re-entrant phase that happens at a higher temperature scale $\sim 50
\mu {\rm K}$, not shown in {}Fig. 1.

\begin{figure}[t]
\epsfysize=6cm
{\centerline{\epsfbox{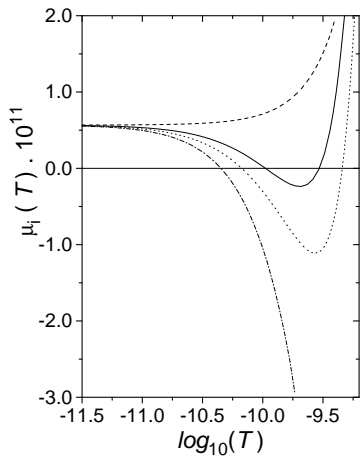}} }
\caption{The effective chemical potentials $\mu_\psi(T)$ 
(dotted and dot-dashed curves, for sets II and III, respectively) and 
$\mu_\phi(T)$
(continuous and dashed curves, for sets II and III, respectively)
as a function of temperature (in units of ${\rm eV}$).}
\label{fig1}
\end{figure}

At the intermediate temperatures the boundness condition $g_\psi(T)
g_\phi (T) > 4 g(T)^2$ does not hold, the overall potential becomes
highly attractive and unbounded from below. Then, suddenly, at the
inversion temperature $T=T_g$ the potential becomes highly repulsive and
both type of atoms leave the ground state at once. Contrary to the first
BEC/normal-gas transition, this second passage to the gas phase is a
discontinuous transition due to the sudden sign inversion of $g(T)$.
This transition is shown in {}Fig. 2 for cases II and III for both
$\psi$ and $\phi$. {}For set II both fields go to the gas phase at a
critical temperature $T_g \sim 40 \mu {\rm K}$, while for set III, $T_g
\sim 340 \mu {\rm K}$. These values should be contrasted to the
ones for the critical temperatures for the uncoupled gases, 
of $T_c^\psi \sim 280 {\rm nK}$, for
both cases and $T_c^\phi
\sim 330 {\rm nK}$ and $\sim 1800 {\rm nK}$, for cases II and III,
respectively.

\begin{figure}[t] 
\epsfysize=6cm  
{\centerline{\epsfbox{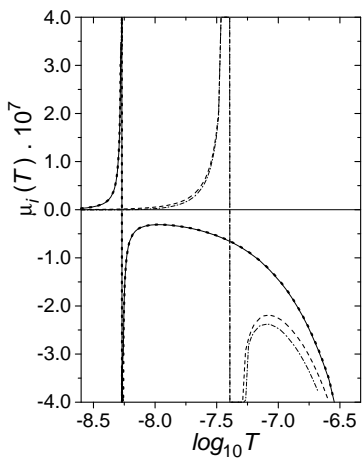}}}
\caption{The high temperature discontinuous transitions suffered by both 
fields. Same notation as in {}Fig. 1.}
\label{fig2}
\end{figure}

Here, we cannot furnish more details
concerning this discontinuous transition since, at the transition,
our high temperature approximation and the effective model of two body interactions
are not valid due to the very large values acquired by the $\mu_i(T)$ and $g(T)$.
However, this should not invalidate the qualitative features of the transition,
at least on its neighborhood.

In conclusion our results show that a system of coupled Bose gases with
an overall repulsive potential, but attractive crossed interactions, may
display unusual transition patterns,
when the important temperature
dependence of the couplings are taken into account. These patterns
include the usual continuous transition condensate/normal-gas followed,
at intermediate temperatures, by an unexpected re-entrant, continuous,
normal-gas/condensate transition. Higher temperatures induce a sudden
change in the sign of the crossed coupling followed by a dramatic
discontinuous transition of the type condensate/normal-gas. Those phases
suggest a possible pattern for the observation of collapsing and
exploding condensates. It is important to note that experimentally one
knows that a change on the sign of the couplings may be achieved by
adjusting external magnetic fields \cite {magn}. However, our results
further indicate that, at least for coupled systems, the temperature can
also act as the external agent which drives the sign inversion. In addition,
our results explicitly demonstrate that, contrary to the relativistic
case, and according to intuition, SNR cannot happen in the 
non-relativistic model of hard core spheres with temperature dependent
couplings. {}Finally, the coupled Bose-Einstein gas system with attractive
crossed interaction also seems to offer an avenue to obtain condensation at
higher temperatures.

The authors  were partially supported by CNPq-Brazil.


\begin{thebibliography}{99}


\bibitem{weinberg} S. Weinberg, Phys. Rev. {\bf D9}, 3320 (1974).

\bibitem{borut} B. Bajc, hep-ph/0002187.

\bibitem{MR1}M. B. Pinto and R. O. Ramos, Phys. Rev {\bf D61}, 125016 (2000).

\bibitem{roos}T. G. Roos, Phys. Rev. {\bf D54}, 2944 (1996).

\bibitem{becreview}F. Dalfovo, S. Giorgini, L. P. Pitaevskii and S. Stringari,
Rev. Mod. Phys. {\bf 71}, 463 (1999);
Ph. W. Courteille, V. S. Bagnato and V. I. Yukalov,
Laser Phys. {\bf 11}, 659 (2001).

\bibitem{myatt} C. J. Myatt {\it et al}., Phys. Rev. Lett. {\bf 78}, 586 (1997);
M. R. Matthews {\it et al}., Phys. Rev. Lett. {\bf 81}, 243 (1998).

\bibitem{coupled} H. Pu and N. P. Bigelow, Phys. Rev. Lett. {\bf 80}, 1130 (1998);
P. Ao and S. T. Chui, Phys. Rev. {\bf A58}, 4836 (1998);
D. M. Jezek and P. Capuzzi, cond-mat/0202025.

\bibitem{norway} T. Haugset, H. Haugerud and F. Ravndal, Ann. Phys. {\bf 226},
27 (1998).

\bibitem{pathria} R. K. Pathria,
{\it Statistical Mechanics} (Pergamon Press, Oxford, 1972).

\bibitem{magn}S. L. Cornish {\it et al}.,
Phys. Rev. Lett. {\bf 85}, 1795 (2000).



\end{thebibliography}
\end{document}